\documentclass[fleqn,usenatbib]{mnras}

\usepackage{newtxtext,newtxmath}
\usepackage[T1]{fontenc}
\usepackage[dvipsnames]{xcolor}

\DeclareRobustCommand{\VAN}[3]{#2}
\let\VANthebibliography\thebibliography
\def\thebibliography{\DeclareRobustCommand{\VAN}[3]{##3}\VANthebibliography}

\usepackage{graphicx}	% Including figure files
\usepackage{amsmath}	% Advanced maths commands

\title[SFR in AGN hosts]{Determining star formation rates in AGN hosts from strong optical emission lines}

\author[De Mellos M. S. Z., et al.]{Maitê S. Z. de Mellos,$^{1,2}$\thanks{E-mail: maite.mellos@acad.ufsm.br} 
Rogemar A. Riffel,$^{1,2}$\thanks{E-mail: rogemar@ufsm.br} 
Jaderson S. Schimoia,$^{1,2}$ 
Sandro B. Rembold,$^{1,2}$ 
\newauthor 
Rogério Riffel,$^{3,2}$ 
Thaisa Storchi-Bergmann,$^{3,2}$ 
Dominika Wylezalek,$^{4}$ 
Gabriele S. Ilha,$^{5,2}$ 
Marco Albán,$^{4}$
\newauthor
Oli L. Dors,$^{6}$ 
Lara Gatto,$^{3,2}$ 
Angela C. Krabbe,$^{4}$ 
Nicolas D. Mallmann,$^{2,3}$ 
Marina Trevisan$^{3}$
\\
$^{1}$Departamento de Física, Centro de Ciências Naturais e Exatas, Universidade Federal de Santa Maria, 97105-900, Santa Maria, RS, Brazil\\
$^{2}$Laboratorio Interinstitucional de e-Astronomia - LIneA, Rua Gal. José Cristino 77, Rio de Janeiro, RJ - 20921-400, Brazil\\
$^{3}$Departamento de Astronomia, Instituto de F\'\i sica, Universidade Federal do Rio Grande do Sul, CP 15051, 91501-970, Porto Alegre, RS, Brazil\\ 
$^{4}$Zentrum für Astronomie der Universität Heidelberg, Astronomisches Rechen-Institut Mönchhofstr, 12-14 69120 Heidelberg, Germany\\
$^{5}$Departamento de Astronomia, Instituto de Astronomia, Geofísica e Ciências Atmosféricas da USP, Cidade Universitária, 05508-900 São Paulo, SP, Brazil\\
$^{6}$ Universidade do Vale do Para\'{\i}ba, Av. Shishima Hifumi, 2911, Zip Code 12244-000, S\~ao Jos\'e dos Campos, SP, Brazil
}

\date{Accepted XXX. Received YYY; in original form ZZZ}

\pubyear{2024}

\begin{document}
\label{firstpage}
\pagerange{\pageref{firstpage}--\pageref{lastpage}}
\maketitle

\begin{abstract}
The influence of Active Galactic Nuclei (AGN) on star formation within their host galaxies remains a topic of intense debate. One of the primary challenges in quantifying the star formation rate (SFR) within AGN hosts arises from the prevalent assumption in most methodologies, which attribute gas excitation to young stars alone. However, this assumption does not consider the contribution of the AGN to the ionization of the gas in their environment. To address this issue, we evaluate the use of strong optical emission lines to obtain the SFR surface density ($\Sigma{{\rm SFR_{AGN}}}$) in regions predominantly ionized by an AGN, using a sample of 293 AGN hosts from the MaNGA survey, with SFR measurements available through stellar population fitting. We propose calibrations involving the H$\alpha$ and [O\,{\sc iii}]$\lambda$5007 emission lines, which can be used to determine $\Sigma{{\rm SFR_{AGN}}}$, resulting in values consistent with those estimated through stellar population fitting. 
\end{abstract}

\begin{keywords}
galaxies: active -- galaxies: star formation -- galaxies: evolution -- galaxies: stellar content
\end{keywords}

%%%%%%%%%%%%%%%%%%%%%%%%%%%%%%%%%%%%%%%%%%%%%%%%%%

%%%%%%%%%%%%%%%%% BODY OF PAPER %%%%%%%%%%%%%%%%%%

\section{Introduction}\label{sec:intro}

%This is a simple template for authors to write new MNRAS papers.
%See \texttt{mnras\_sample.tex} for a more complex example, and \texttt{mnras\_guide.tex}
%for a full user guide.

%All papers should start with an Introduction section, which sets the work
%in context, cites relevant earlier studies in the field by \citet{Fournier1901},
%and describes the problem the authors aim to solve \citep[e.g.][]{vanDijk1902}.
%Multiple citations can be joined in a simple way like \citet{deLaguarde1903, delaGuarde1904}.

It is consensus that all galaxies with stellar bulges host a Supermassive Black Hole (SMBH) in their center. However, only $\sim10\%$ of these galaxies exhibit an energetic phenomenon in their core \citep{ho2008}, which manifests only in the presence of a gas reservoir to fuel the SMBH. This compact region is known as an Active Galactic Nucleus (AGN) and harbors a SMBH with masses of 10$^6-$10$^{10}$M$_{\odot}$. 
%The matter captured due to the gravitational field tends to form an accretion disk around the SMBH, generating heat through the transformation of gravitational potential energy into kinetic energy and electromagnetic radiation \citep{urry1995}.
Over the decades, the relation between the AGN and its host galaxy has been explored, going back to initial ideas about the connection between the dense nuclear gas concentration observed in infrared (IR) galaxies and the feeding of the SMBH \citep{joseph, sanders}. Several observations have shown the existence of a correlation between the SMBH mass and its host galaxy properties, such as stellar velocity dispersion  \citep{ferrarese, bulgeGebh}, luminosity and stellar mass \citep{bulgemass, magorrian1998, nadine2004}, and the evolution of star formation \citep{coevolution_review}, support the conclusion of a co-evolution between the AGN and its host galaxy. The processes of AGN feeding and feedback are the principal regulatory agents in this relation \citep[e.g.][]{harrison_review, harrison24, thaisa&allan2019}. 

The feeding process plays an important role in the evolution of the host galaxy. Since the AGN luminosity is an indicator of nuclear activity, the luminosity varies with the availability of gas: low-luminosity AGNs may be fed with gas from the mass loss of evolving stars in the bulge, while more luminous AGNs require an additional source of gas in a short time interval \citep{thaisa&allan2019}. In contrast, AGN feedback, manifested in the form of radiation, jets and winds powered by the accretion disk, can heat and expel gas from the nuclear region, suppressing stellar formation \citep{fabian2012, Symeonidis2012} and regulating galaxy growth, preventing the formation of excessively massive systems \citep{harrison_review}. A significant impact in this regard is observed when measuring the molecular mass outflow rates in luminous galaxies in the local universe. In many cases, this rate exceeds the competing star formation rate (SFR), meaning that the material conducive to forming stars is being removed more rapidly than the galaxy is capable of forming new stars \citep{molecularoutflows}. Observational studies also reveal the presence of regions in the same galaxy that simultaneously exhibit more and less intense star formation. This difference is attributed to the influence of AGN outflows. This phenomenon creates an environment where the conditions for forming new stars vary drastically, leading to regions experiencing a star formation burst, while others undergo suppression \citep{negandposfeedback,Quasargalaxyformation}.

Moreover, numerical simulations demonstrate that AGN feedback can also have positive effects regarding star formation. In galactic nuclei with a high molecular content, AGN outflows exhibit the capacity to intensify star formation by compressing dense cold gas, leading to star forming burst. This dynamic establishes a relation between the internal pressure of the AGN outflow and the triggering of star formation in the galaxy disk, suggesting that appropriate internal pressure could have beneficial effects on star formation, compressing the gas and stimulating the process \citep{zubovas2013}. Observational studies, such as those conducted by \cite{rogerio2023}, indicate that in galaxies with a more luminous AGN, the contribution of the young stellar population appears to increase as the galactocentric distance decreases, i.e, closer to the AGN. This suggests that this behavior is due to the availability of gas in this region, originating from the mass loss of stars of intermediate age, which results in an additional fuel supply to the AGN.

Thus, there is evidence that the AGN can exert influence over star formation in its host galaxy \citep{rogerio2024}, enhancing the formation of new stars in some cases and quenching it in others. The impact of the AGN on star formation in the local universe is still widely debated.  However, unlike in star-forming galaxies, measuring the star formation rate in AGN hosts can be quite challenging. For non-active galaxies, there are many calibrators to determine the SFR as it scales with the gas density, a relation known as the Kennicutt–Schmidt law \citep{schmidt59,kennicutt89,kennicutt1998,Kennicutt12}. In active galaxies, the AGN emission can "contaminate" many of the SFR indicators, such as the H$\alpha$ luminosity, which is enhanced due to the emission of gas being photoionized by the AGN, rather than solely by young stars. In infrared-selected AGN hosts, the AGN contribution to the total emission is related to the galaxy luminosity function, rising as the luminosity increases. Moreover, this contribution also depends on redshift, with AGN contributions becoming more prominent at higher redshifts \citep{Symeonidis10,Symeonidis2018, Symeonidis2019, Symeonidis2021, Symeonidis2022,herman-caballero11,Diaz-Santos17}.

In the optical, the star formation rates in AGN hosts can be determined through stellar population spectral synthesis  \citep{rogerio2021,rogerio2023}. This method typically requires spectra with high signal-to-noise ratio (S/N \ $\gtrsim$ \ 10) and are time-demanding procedure. In the present study, we use a sample of 293 AGN hosts, selected as described in \cite{sandro2017} and \cite{rogerio2023}, from the Mapping Nearby Galaxies at Apache Point Observatory (MaNGA; \citeauthor{bundy2015} \citeyear{bundy2015}) survey, to explore the relation between the optical emission line luminosities and the SFR, using the measurements described in \citet{rogerio2023}.

This paper is structured as follows. In Section \ref{sec:data}, we describe our sample and the data used in this work, while in Section \ref{sec:sfr} we describe the methodology used to derive the SFRs. The results and discussions are presented in Section \ref{sec:res_dis} and \ref{sec:disc}. Finally, in Section \ref{sec:conclusion} we present our conclusions. Throughout the paper, we used $H_0$ = 73 km s$^{-1}$ Mpc$^{-1}$ \citep{hubbleconstant}.

\section{The Data} \label{sec:data}
One of the most significant observational projects to date is the Sloan Digital Sky Survey (SDSS), which was designed to map the sky by collecting optical spectroscopic and photometric data (3600 - 10,400 \AA) using the Sloan Foundation Telescope \citep{sdss}. Utilizing Integral Field Spectroscopy (IFS), the MaNGA \citep{bundy2015} survey is one of the three principal projects of the fourth generation of the SDSS. 

Its main goal was to investigate the kinematic structure and gas composition of $\sim$ 10,000 galaxies in the local universe ($\langle z \rangle \approx 0.03$). The MaNGA Integral Field Units (IFU) are composed of a set of optical fibers with their sizes varying according to the number of these fibers \citep{drory2015, law2015, yan2016, yan2016b}. This variation generates different fields of view, ranging from 12$''$ to 32$''$.

We used the Seventeenth SDSS Data Release (DR17) data cubes, which contain 293 active galaxies \citep{rogerio2023}, to select only spaxels that exhibit AGN ionization. The optical classification of AGN hosts is based on the ionizing source of the gas by utilizing the [O\,{\sc iii}]$\lambda$5007/H$\beta$ versus [N\,{\sc ii}]$\lambda$6584/H$\alpha$ diagnostic diagram proposed by \cite{bpt}, commonly known as BPT diagram, and the diagram introduced by \cite{whan}, which uses the equivalent width of H$\alpha$ (W$_{\rm H \alpha}$) and is known as WHAN diagram, as outlined by \cite{sandro2017}. Consequently, a galaxy is confirmed as an optical AGN host when it is positioned simultaneously in the Seyfert/LINER region on both the BPT and WHAN diagrams (see \citeauthor{sandro2017} \citeyear{sandro2017} for more details). According to the Galaxy Zoo Project \citep{galaxyzoo}, 32\%, 60\%, and 5\% of our sample are elliptical, spiral, and merging galaxies, respectively.

\section{Methodology} \label{sec:sfr}

\subsection{Determining the SFR from stellar population synthesis}

With the aim of interpreting galaxy spectra through the superposition of stellar spectra, stellar population fitting is an invaluable tool for comprehending the formation and evolution of galaxies. The stellar population synthesis provides essential information about stellar age distributions, metallicity, extinction, velocity dispersion, stellar mass, and stellar formation history by fitting the spectra with a linear combination of $N_{\star}$ simple stellar populations (SSP), which is calculated from the evolutionary synthesis models. 

Here, we use the SFRs derived in \citet{rogerio2023}, by using the {\sc starlight} code \citep{starlight2005}. The base of elements we used is the GM described in \citet{cid2013, cid2014}, which is constructed using the {\sc Miles} \citep{vazdekis2010} and \citet{gonzalez2005} models, with a power law of the form $F_\nu \propto \nu^{-1.5}$. 

 The presence of extinction caused by dust in observed galaxies is parametrized by the extinction $A_V$ in V band, adopting the galactic extinction law of \cite{cardelli1989} with $R_V = 3.1$. Additionally, the stellar motion along the line of sight is modeled by a Gaussian distribution $G$ centered on the velocity $v_{\star}$ with dispersion $\sigma_{\star}$. Therefore, the model spectrum is given by 
\begin{equation}
    M_\lambda = M_{\lambda_0} \left[ \sum_{j=1}^{N_\star} x_j b_{j,\lambda} r_\lambda \right] \otimes G(v_\star, \sigma_\star),
\end{equation} where $b_{j,\lambda}$ is the spectrum of the j-th SSP normalized at the wavelength $\lambda_0$, $r_\lambda \equiv 10^{-0.4(A_\lambda - A_{\lambda_0})}$ represents the reddening term, $M_{\lambda_0}$ is the synthetic flux at the normalization wavelength, $\vec{x} = (x_1,...,x_{N_\star})$ is the population vector, and $\otimes$ denotes the convolution operator. The Markov Chain Monte Carlo method is employed for parameter fitting, seeking a minimum in the parameter space based on $\chi^2 = \Sigma_\lambda [(O_\lambda - M_\lambda)\omega_\lambda]^2$, where $ O_{\lambda}$ is the observed spectrum and $M_{\lambda} $ its corresponding model. Throughout this process, certain restrictions are applied, such as setting a weight $\omega_\lambda = 0$ in regions containing emission lines.

Furthermore, the use of {\sc starlight} makes it possible to derive the star formation rate across a range of stellar population ages, denoted by $\Delta t = t_{j_f} - t_{j_i}$ \citep{Asari+07,rogerio2021}. This assumes that the mass of each component of the base transformed into stars, measured in $M_{\odot}$, is given by

\begin{equation}
    M_{\star, j}^{ini} = \mu_{j}^{ini} \times \frac{4 \pi d^2}{ L_{\odot}},
\end{equation} where $ \mu_{j}^{ini}$ represents the flux corresponding to the mass converted into stars for the j-th element, given in M$_{\odot}$ erg s$^{-1}$ cm$^{-2}$, $d$ is the distance of the galaxy in centimeters, and $L_{\odot}$ is the solar luminosity in $\rm erg \ s^{-1}$. Thus, the star formation rate in a time interval will be

\begin{equation}
    \text{SFR}_{\star} = \frac{\Sigma_{j_i}^{j_f} M_{\star,j}^{ini}}{\Delta t}.
\end{equation}

In this work, we use the SFR$_{\star}$ measurements obtained by \cite{rogerio2023} for the time interval $\Delta t = 20 $ Myr, which is shown to closely reproduce the H$\alpha$ based measurements of the star formation rate (SFR$_{\rm R21}$) in regions dominated by the ionization of young stars \cite{rogerio2021}. We followed the same procedure described by these authors to recalculate the calibration between the gas and stellar based SFR densities. This recalibration is justified by the fact that we use the updated sample from \cite{rogerio2023} and use a different Hubble constant than that used in \cite{rogerio2021}. We only use SF dominated spaxels from the control sample, excluding spaxels where the stellar population synthesis includes a contribution of the featureless continuum (FC) and spaxels with H$\alpha$ equivalent width smaller than 3 \AA. This results in the following relation 
%In short, these authors found that the SFR density ${\Sigma{\text{SFR}_{\text{R21}}}}$ can be estimated by 

%\begin{equation}\label{eq_riffel}
%    \log({\Sigma{\text{SFR}_{\text{R21}}}}) = (1.045 \pm 0.006) %\log({\Sigma{\text{SFR}_{\star}}}) - (0.080 \pm 0.009),
%\end{equation}
%obtained by using SF dominated spaxels in AGN hosts, with the SFR density given in units of  M$_{\odot}$ yr$^{-1}$ kpc$^{-2}$. 

%\textcolor{red}{*** atualizar esse texto *** usando somente spaxels SF das controles, sem contribuição de FC, a equação de calibração é}

\begin{equation}\label{eq_riffel}
    \log({\Sigma{\text{SFR}_{\text{R21}}}}) = (1.170 \pm 0.007) \log({\Sigma{\text{SFR}_{\star}}}) - (0.143 \pm 0.014),
\end{equation} where ${\Sigma{\text{SFR}_{\text{R21}}}}$ is the gas based SFR density estimated from the stellar population values using the procedure described in \citet{rogerio2021},  with the SFR density given in units of  M$_{\odot}$ yr$^{-1}$ kpc$^{-2}$. 

%\textcolor{red}{obtained by using SF dominated spaxels in AGN hosts, with the SFR density given in units of  M$_{\odot}$ yr$^{-1}$ kpc$^{-2}$. }

%Due to the substantial capacity of the AGN to ionize the gas, determining the SFR in these active galaxies using conventional methods, which assume that the ionized photons originate only from young stars, is a challenging task. A method to evaluate two approaches for determining the SFR in regions dominated only by the ionization of young stars was proposed by \cite{rogerio2021}. In this study, the authors selected the star-forming spaxels of a sample composed of 170 active galaxies combined with 291 control galaxies from Survey MaNGA, with the aim of comparing the SFR from the H$\alpha$ line luminosity, according to the \cite{kennicutt1998} relation, with the SFR obtained by the stellar population synthesis carried out by STARLIGHT, for each data cube spaxel, considering stellar populations of different ages. However, the best result was attained for the stellar population with an maximum age of 20 Myr. The established relation is

%\begin{equation}\label{eq_riffel}
%    \log({\Sigma{\text{SFR}_{\text{R21}}}}) = (1.147 \pm 0.005) \log({\Sigma{\text{SFR}_{\star}}}) + (0.086 \pm 0.080).
%\end{equation}

\begin{figure}
 \includegraphics[width=\columnwidth]{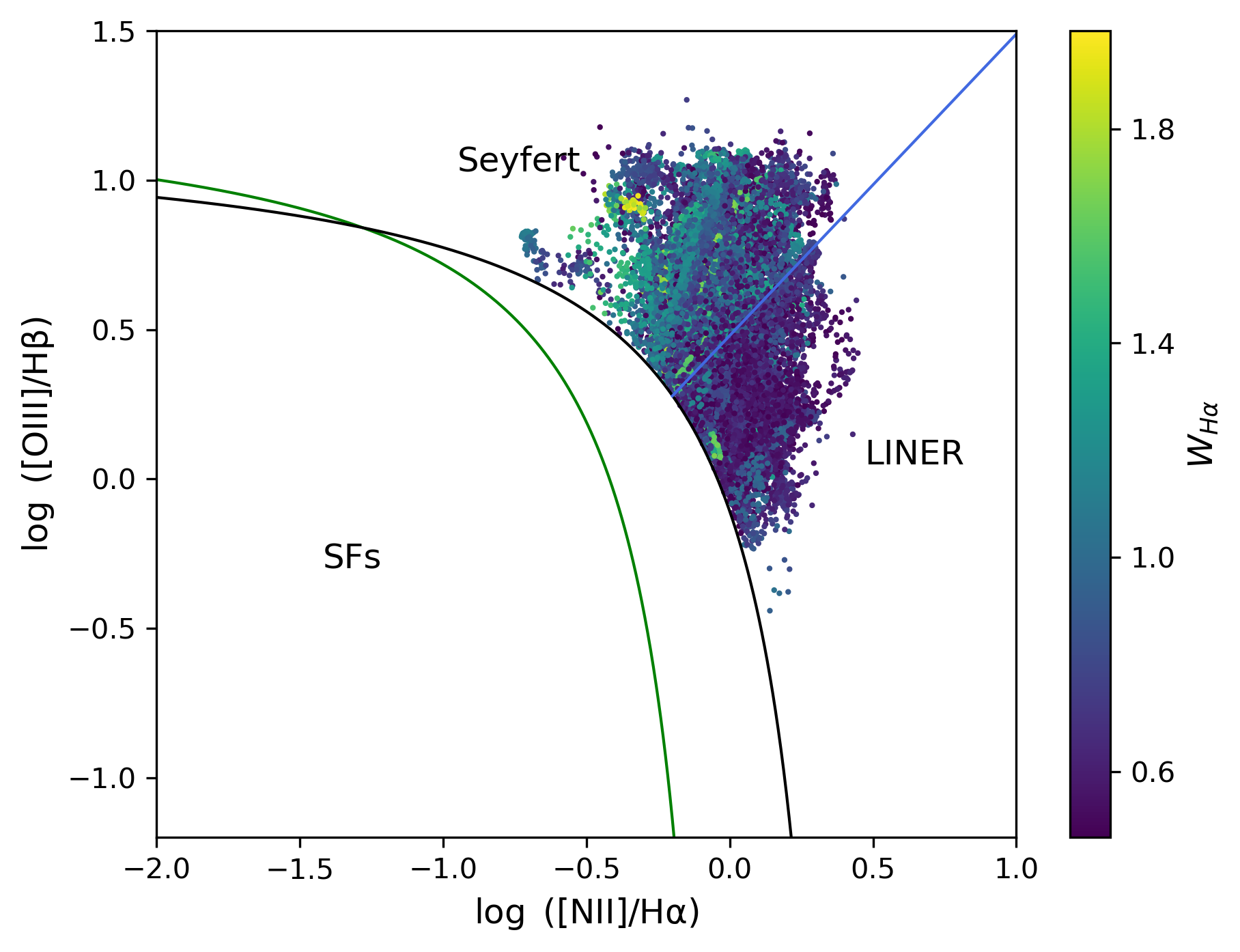}
 \caption{BPT diagnostic diagram of the 22,084 selected spaxels from a sample of 293 active galaxies. These spaxels exhibit high-quality data, a well-fitted stellar population synthesis, and predominantly AGN-ionized gas as discussed in Sec. \ref{subsec:spaxels_sel}. The black curve represents the theoretical upper limit for the star-forming regions (SFs) proposed by \citet{kewley2001}, the green curve is the empirical star-forming limit proposed by \citet{kauffmann2003}, and the blue line represents the separation between Seyferts and LINERs from \citet{kewley2006}. The region between the green and black lines is denominated the composite region. } 
 \label{fig:bpt_spaxel}
\end{figure}

\begin{figure}
    \centering
\includegraphics[width=0.48\textwidth]{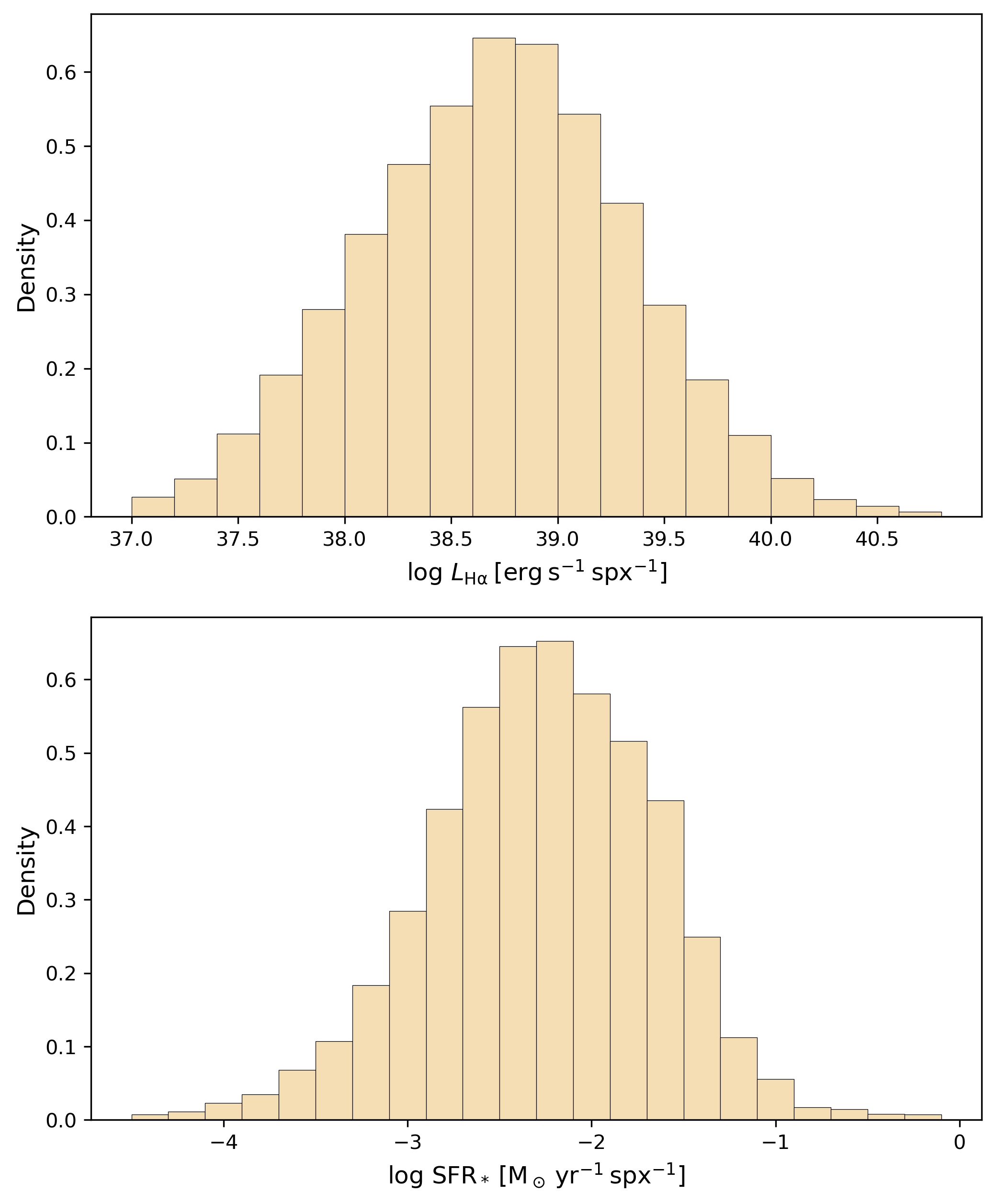}
    \caption{Density histograms for all 22,084 spaxels from 293 active galaxies predominantly ionized by AGN. The top panel represents the density histogram for $L_{\rm H \alpha}$, while the bottom panel displays the density histogram for $\text{SFR}_{\star}$.} 
    \label{fig:histogram}
\end{figure}

\subsection{Spaxel selection} \label{subsec:spaxels_sel}

The emission line fluxes were obtained from \citet{rogerio2023}, and are based on the fitting of Gaussians to the line profiles using the {\sc ifscube} python package \citep{ifscube}. In order to choose only high quality spaxels predominantly displaying AGN-ionized gas, we employ the following criteria:

\begin{itemize}
\item [(i)] For a spaxel to be identified as predominantly ionized by an AGN, it is necessary to attend the following conditions:
        \begin{itemize}
            \item[a.] $\log(\text{[O\:{\sc iii}]}/{\rm H}\beta) \geq \frac{0.61}{\log({\rm [NII]/H}\alpha) - 0.47} + 1.19$ \citep{kewley2001};
            \item [b.] $\log(W_{\rm H\alpha}) \geq 0.5 \AA $ \citep{whan};

            \item [c.] $\log({\rm [NII]/H}\alpha) > -0.4$ \citep{whan}.
        \end{itemize}

\item[(ii)] It is necessary that the mean S/N of each spaxel in the continuum window between 5650 and 6750 \AA\ is S/N $\geq$ 10. In addition, we exclude spaxels with adev=$| O_{\lambda} - M_{\lambda} | / O_{\lambda} < 10\%$. This ensures the reliability of the stellar population fit performed by {\sc starlight} \citep{starlight2005}.

%\item[(iii)] The adev(\%) of the corresponding spaxel must be $| O_{\lambda} - M_{\lambda} | / O_{\lambda} < 10$ ;

%\item[(iv)] The extinction in the V-band, denoted by $A_v$, must be in the range $0 < A_v < 4$.
\end{itemize}

Hence, by using these criteria, the selected spaxel exhibits high-quality data, a well-fitted stellar population, and predominantly AGN ionized gas. The total number of selected spaxels from all AGN hosts is 22,084, with the BPT diagram represented in Figure \ref{fig:bpt_spaxel}. According to the BPT diagram, 55.32\:\% of the spaxels lie in the Seyfert region, while the remaining are in the LINER region.

%\subsection{SFR from H$\alpha$ emission-line fluxes}

%Due to the substantial capacity of the AGN to ionize the gas, determining the SFR in these active galaxies using conventional methods, which assume that the ionized photons originate only from young stars, is a challenging task. A method to evaluate two approaches for determining the SFR in regions dominated only by the ionization of young stars was proposed by \cite{rogerio2021}. In this study, the authors selected the star-forming spaxels of a sample composed of 170 active galaxies combined with 291 control galaxies from Survey MaNGA, with the aim of comparing the SFR from the H$\alpha$ line luminosity, according to the \cite{kennicutt1998} relation, with the SFR obtained by the stellar population synthesis carried out by STARLIGHT, for each data cube spaxel, considering stellar populations of different ages. However, the best result was attained for the stellar population with an maximum age of 20 Myr. The established relation is

%\begin{equation}\label{eq_riffel}
%    \log({\Sigma{\text{SFR}_{\text{R21}}}}) = (1.147 \pm 0.005) \log({\Sigma{\text{SFR}_{\star}}}) + (0.086 \pm 0.080).
%\end{equation}

%From this equation, it is possible to calculate the star formation rate through stellar population synthesis, in a star-forming region of a galaxy. This is obtained in units of M$_{\odot}$ yr$^{-1}$ kpc$^{-2}$. 

\subsection{Correction of the interstellar extinction}\label{subsec:reddening}

The scattering and the absorption of light from luminous sources by the interstellar dust results in reddening and reduces the amount of the observed light. Therefore, it is necessary to correct the obtained spectra by this effect. The emission-line fluxes are corrected for extinction by using
%To achieve this, we employ the equation
\begin{equation}
    F_{\rm int} = F_{\rm obs} 10^{0.4\:A_{\lambda}},
\end{equation} with $F_{\rm int}$ being the intrinsic flux, $F_{\rm obs}$ the observed flux, and $A_{\lambda}$ is the interstellar extinction at the wavelength $\lambda$, given by 
\begin{equation}
    A_{\lambda} = A_{\rm v} q_{\lambda},
\end{equation}  
where the extinction in the V band can be estimated from the observed ratio between the H$\alpha$ and H$\beta$ fluxes, $(F_{\rm H\alpha}/F_{\rm H\beta})_{\rm obs}$, using
\begin{equation}
    A_{\rm v} = 7.22 \log \left( \frac{(F_{\rm H\alpha}/F_{\rm H\beta})_{\rm obs} }{2.86}  \right),
\end{equation}
where it was adopted a theoretical intensity line ratio of $F_{\rm H\alpha}/F_{\rm H\beta}=2.86$ for case B \ion{H}{i} recombination at the low-density limit and electron temperature of $T_{\rm e} = 10^4$ K  \citep{osterbrock}.

We use this procedure, together with the extinction law from \citet{cardelli1989}, to obtain the extinction-corrected emission line luminosities in all spaxels. These corrected luminosities are then used to investigate the relation between the gas emission and star formation rate in AGN dominated regions.

%We used this procedure to correct the observed flux of the H$\alpha$ and [O\:{\sc iii}]$\lambda$5007 emission line, which was obtained utilizing the IFSCUBE package \citep{ifscube}, in order to determine L(H$\alpha$) and L([O\:{\sc iii}]) in all spaxels exhibiting AGN ionization. We then use $q_{\lambda} = 0.818$ and $q_{\lambda} = 1.12$, which corresponds to the value for H$\alpha$ and [O\:{\sc iii}]$\lambda$5007, respectively, with $R = 3.1$ \citep{cardelli1989}.

\subsection{The effect of the featureless continuum in the determination of SFR}\label{subsec:fc}

An additional challenge arises when analyzing the stellar populations of AGN galaxies due to the degeneracy between a young stellar population and a featureless continuum (FC). Essentially, it is very difficult to distinguish a reddened young starburst from a FC continuum in the optical region of the spectrum, because the main differences between a $\sim$5 Myr SSP and an FC seen through dust, with an absorption $A_v \sim$ 2–3 mag, is the presence of the Balmer absorption lines and
Balmer jump in the blue side of the electromagnetic spectrum in the former \citep{cid2004,rogerio2009}. Thus, while a significant fraction of AGN spectra may require a FC component, this does not necessarily indicate the presence of an AGN, the continuum may mimic the presence of young stellar populations \citep[e.g.][]{CidFernandes+95, CidFernandes+98,cid2004,Vega+09,rogerio2009,Bon+14,Cardoso+17,rogerio2022,rogerio2024}.

For instance, \citet{CidFernandes+92,CidFernandes+95} have shown that H$\beta$ line only shows a broad component in polarized light when FC$\gtrsim$ 20\%, thus values of FC below this are degenerated and can be attributed both to an AGN or SF components. In our analysis, for 62.3\,\% of the spaxels, no FC component is required to fit their spectra, while FC contributions lower than 10\,\% are found for 32.2\,\% of the spaxels and FC contributions above  10\,\%  are found in only 5.5\,\% of the spaxels. The mean $\log \Sigma{\text{SFR}_{\star}}$ and its standard deviation within these groups are $-1.59\pm0.53$, $-1.60\pm0.64$ and $-1.24\pm0.53$, respectively, given in units of M$_{\odot}$ yr$^{-1}$ kpc$^{-2}$. Thus, if the FC component is masking the contribution of a young stellar population, the introduced bias is at most 0.36 dex, in our sample. This value is much lower than the observed scatter in star formation rate density for different H$\alpha$ luminosities, as reported by \citet{rogerio2021}. With this in mind, we can estimate the star formation densities for our AGN sample using Eq.~\ref{eq_riffel} and investigate possible relations with their emission emission line properties.

\section{Results} \label{sec:res_dis}

Fig. ~\ref{fig:histogram} shows the distribution of the extinction corrected H$\alpha$ luminosities ($L_{H\alpha}$) in the top panel and the SFR$_{\star}$ over the last 20 Myr from \citet{rogerio2021} in the bottom panel. As shown in this figure, the selected spaxels span a range of four orders of magnitude in $L_{H\alpha}$ and a similar range in star formation rates. 

\begin{figure*}
\includegraphics[width=0.9\textwidth]{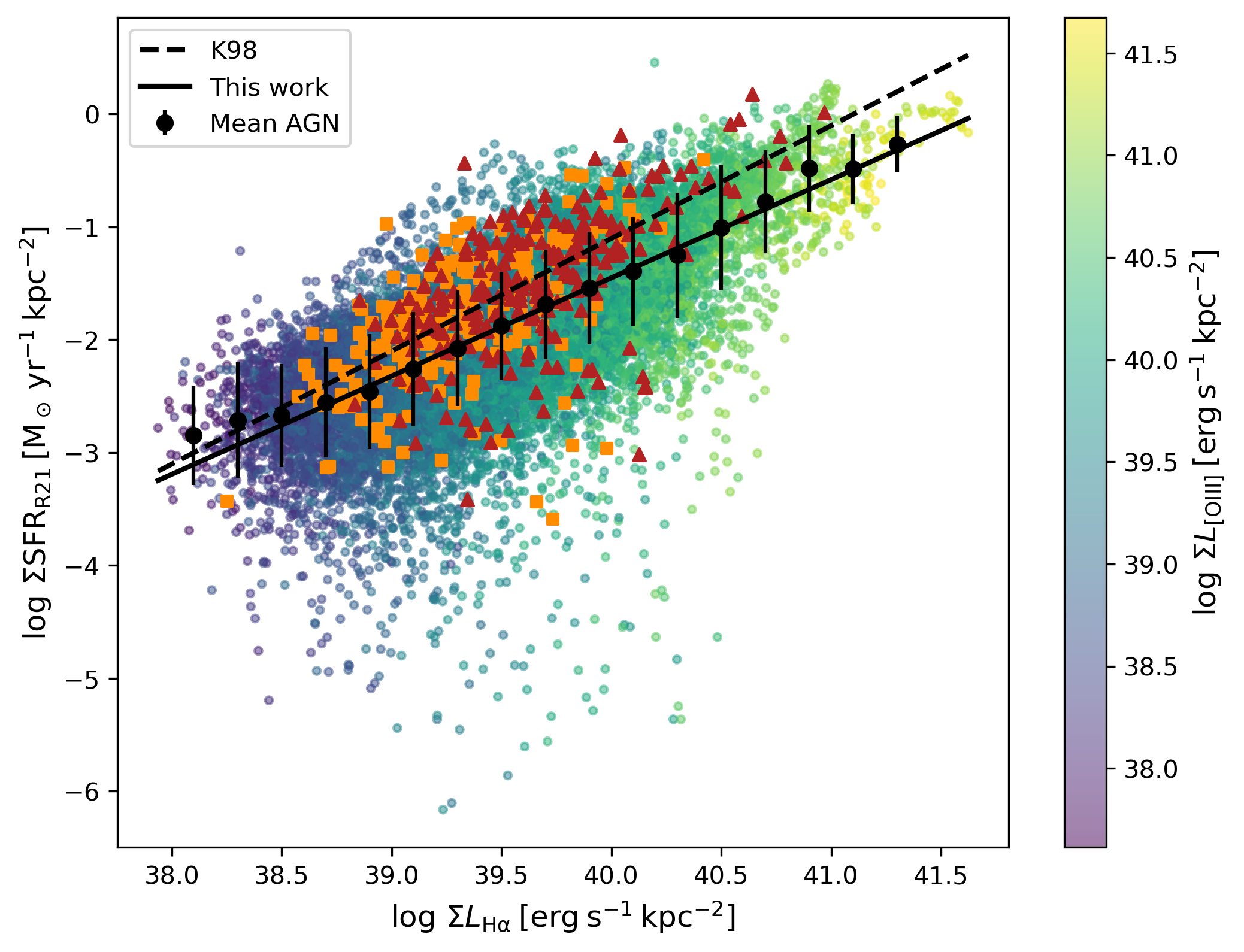}
\caption{Relation between $\Sigma{\text{SFR}_{\text{R21}}}$ and $\Sigma{\rm L_{H\alpha}}$ involving 22,084 spaxels of 293 active galaxies with predominant AGN-type ionization, identified by their respective values of L([O\:{\sc iii}]). The red triangles and orange squares represent measurements within integrated apertures of 2.5 arcsec and 4 kpc diameter, respectively. The best-fit linear relation is represented in the solid line, while the black circles represent the mean values with their standard deviations of 20 linearly spaced bins of $\Sigma{\rm L_{H\alpha}}$. The dashed line corresponds to the \protect\cite{kennicutt1998} relation.} 
\label{fig:ajustereta}
\end{figure*}

%Assuming the applicability of the Eq.~\ref{eq_riffel} to spaxels predominantly influenced by AGN ionization, we utilize $\Sigma{\text{SFR}_{\star}}$ derived from the stellar population synthesis, to calculate $\Sigma{\text{SFR}_{\rm R21}}$ for the 22,084 selected spaxels, in units of M$_{\odot}$ yr$^{-1}$ kpc$^{-2}$. Using the H$\alpha$ emission line flux obtained by IFSCUBE, which was corrected for reddening (see Section \ref{subsec:reddening}), we obtain the H$\alpha$ emission line luminosity, L(H$\alpha$). 

Dividing the $L_{H\alpha}$ by the respective spaxel area, we obtain H$\alpha$ surface brightness ($\Sigma{\rm L_{H\alpha}}$), in units of erg s$^{-1}$ kpc$^{-2}$, which can be compared to the SFR density. Fig. ~\ref{fig:ajustereta} shows a plot of $\Sigma{\text{SFR}_{\rm R21}}$, obtained from Eq.~\ref{eq_riffel}, versus $\Sigma{\rm L_{H\alpha}}$ considering all the selected spaxels in the AGN dominated region, color-coded by their [O\,{\sc iii}]$\lambda$5007 surface brightness. A strong correlation between $\Sigma{\rm L_{H\alpha}}$ and $\Sigma{\text{SFR}_{\rm R21}}$ is observed, with a Pearson correlation coefficient of $\rho=0.72$. The black circles represent the mean values with their standard deviations within 20 equally spaced bins of $\log\:\Sigma{\rm L_{H\alpha}}$. For comparison, the dashed line shows the values predicted by the \citet{kennicutt1998} relation, if it were applied to the observed range of luminosities. A larger spread of points is observed in spaxels with lower $\Sigma{\text{SFR}_{\rm R21}}$ (below the mean values), i.e. much lower values of $\Sigma{\text{SFR}_{\rm R21}}$ are observed than expected from the \citet{kennicutt1998} relation for a given $\Sigma{\rm L_{H\alpha}}$.  The best linear fit to the data is represented in the black solid line and is given by
\begin{equation}\label{ajuste_total}
   \log (\Sigma{\text{SFR}_{\rm AGN,H\alpha}}) = (0.870 \pm 0.006)\log(\Sigma{\rm L_{H\alpha}}) - (36.273 \pm 0.254),
\end{equation} which can be used to determine the SFR densities in regions dominated by AGN ionization, based solely on the luminosity of H$\alpha$. This equation is parametrized using $\Sigma{\rm L_{H\alpha}}$ in units of erg s$^{-1}$ kpc$^{-2}$ and results in the SFR density in units of M$_{\odot}$ yr$^{-1}$ kpc$^{-2}$. 

Our sample includes galaxies with redshifts in the range $0.016\leq z\leq 0.145$, with a median value of $z=0.043$. Thus, the MaNGA spaxel width of 0.5 arcsec corresponds to physical scales from $\sim$160 pc to $\sim1.51$ kpc, with a median value of $\sim450$ pc. The detection of AGNs using optical emission-line ratio diagnostics is influenced by the aperture size. As the aperture size increases, the rate of AGN detection tends to decrease because contamination from extra-nuclear star-forming regions and diffuse ionized gas becomes more significant  \citep[e.g.][]{Alban23}. To investigate how our results change for integrated spectra, we have included in Fig.\:\ref{fig:ajustereta} the integrated measurements in two apertures: (i) 2.5 arcsec and (ii) 4 kpc in diameter, centered on the position of the continuum peak. The first aperture correspond to the angular resolution of MaNGA datacubes \citep{yan2016c,law2016}, while the 4 kpc aperture was optimized to include the maximum number of objects possible. For the closest objects in our sample, this aperture approximately matches the size of the MaNGA field of view, while for the most distant objects this aperture is just marginally resolved by MaNGA. The MaNGA angular resolution corresponds to physical scales of $\sim830$ pc to $\sim7.53$ kpc at the galaxies of our sample. In our analysis based on the  4 kpc aperture, we exclude objects for which the MaNGA datacube covers a smaller region (1 galaxy) and those where 4 kpc is not resolved by the observations (25 objects).

We integrate the emission-line fluxes within the specified apertures and then select the AGN-dominated integrations. We find that 256 galaxies located in the AGN region of the BPT diagnostic diagram and have $\log\:(W_{\rm H\alpha}) \geq 0.5 \AA$  for the 2.5 arcsec aperture and 238 for the 4 kpc aperture. The resulting points in the plot of $\Sigma{\text{SFR}_{\rm R21}}$ vs. $\Sigma{\rm L_{H\alpha}}$ for the 2.5 arcsec and for the 4 kpc  apertures are shown in  Fig.~\ref{fig:ajustereta} as red triangles and orange squares, respectively. We observe that the points obtained from the integrated spectra fall into the same region as the distribution of points from the individual spaxels. This is somewhat expected, as the properties shown on both axes are normalized by the aperture area. Thus, the relationship obtained above can also be applied to measurements from integrated spectra within these apertures, provided they are classified as AGN-dominated objects. Additionally, we observe a decrease in the number of points at both extremes of the plot, which was anticipated due to the contamination from extra-nuclear emission.  At the high end, the decrease is due to the inclusion of spaxels that are less luminous compared to the nuclear spaxel, reducing the H$\alpha$ surface brightness and the star formation density. At the lower end, objects where some less luminous spaxels were classified as AGN shift to the region occupied by transition objects and star-forming regions in the BPT diagram due to contamination from emission in star-forming regions. For a comprehensive discussion on the impact of aperture size on the selection of AGNs using optical diagnostic diagrams in the MaNGA survey, see \citet{Alban23}. 

 % \textcolor{red}{We strongly don't recommend using this method for integrated spectra across the entire galaxy, as many galaxies are no longer classified as AGNs according to the criteria of the BPT and WHAN diagrams. This is because different regions exhibit different ionization sources, and the highest ionization from the AGN is typically concentrated in the nuclear region.}
 
% \textcolor{red}{In Fig.~\ref{fig:ajustereta} we also show the relation between $\Sigma{\text{SFR}_{\rm R21}}$ and $\Sigma{\rm L_{H\alpha}}$, considering 261 galaxies located in the AGN region of the BPT diagnostic diagram, using a circular extraction with a diameter of 2.5 arcsec, equivalent to the MaNGA PSF. In this case, the SFR of each galaxy was divided by the physical circular area corresponding to the extraction angle, in kpc$^2$. As a result, the points shift downwards and to the left, reducing the number of points the upper right corner. Nevertheless, the points obtained from the circular extraction remain in the same region as the individual spaxels, allowing our calibration to be applied for estimating SFRs from any type of observation, without the need for spatially resolved spectra. }

\begin{figure*}
 \includegraphics[width=0.48\textwidth]{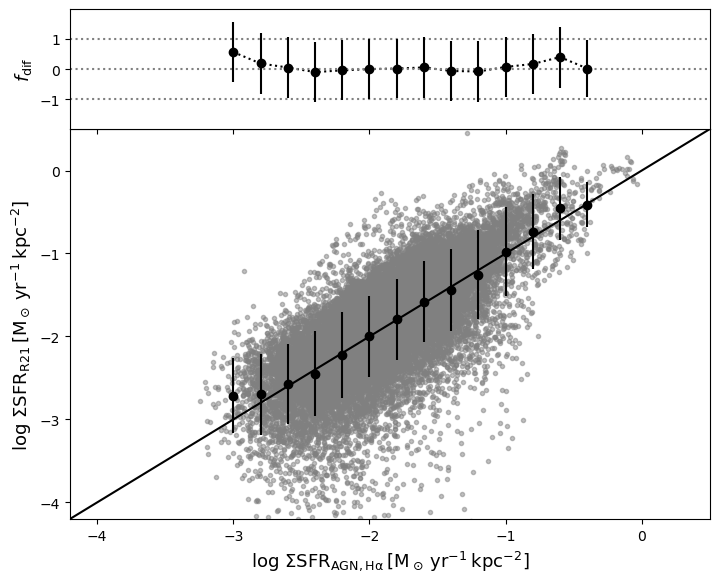}
  \includegraphics[width=0.48\textwidth]{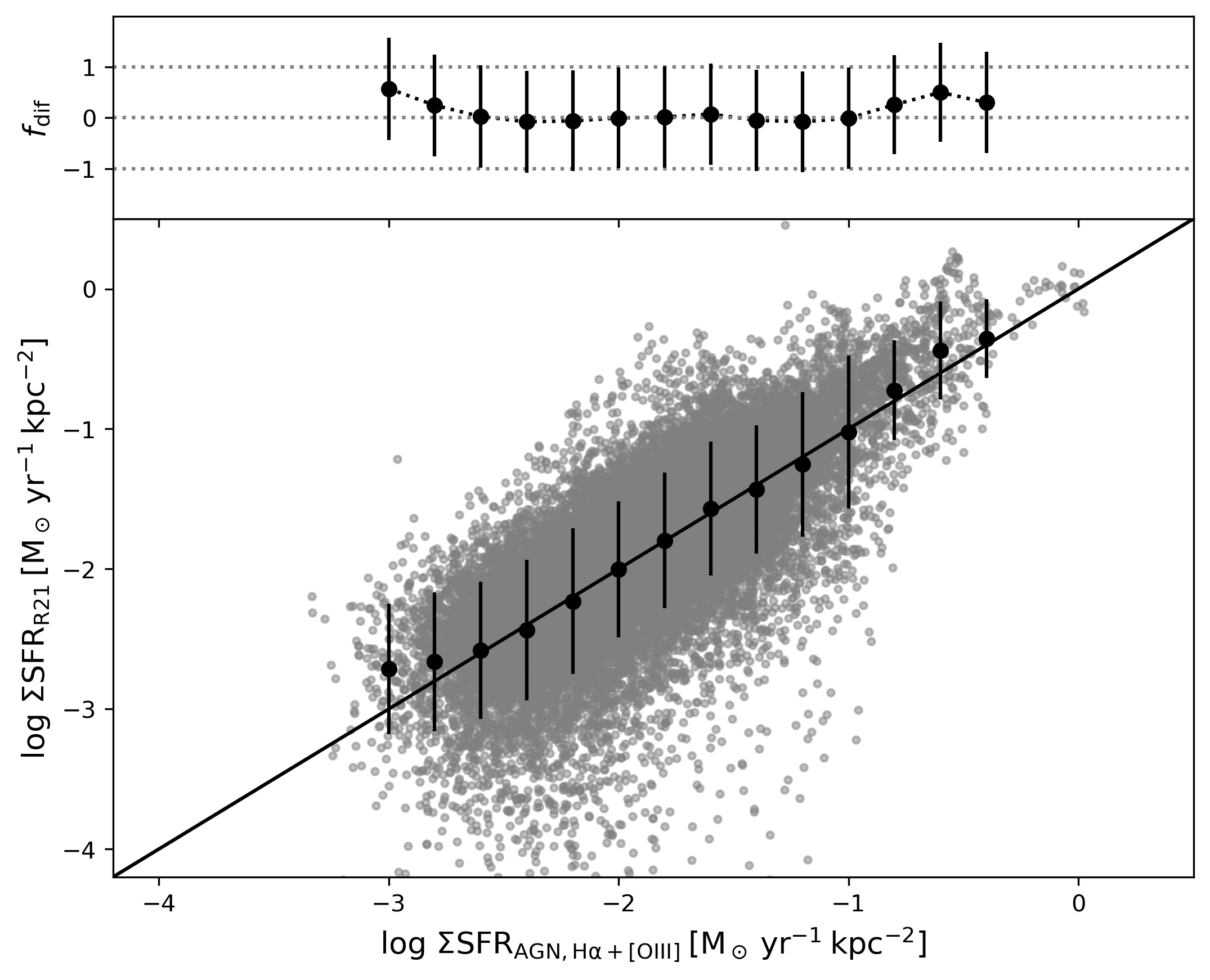}
   \caption{The bottom panels show a comparison of $\Sigma{\text{SFR}_{\text{R21}}}$ with $\Sigma{{\rm SFR_{AGN, H\alpha}}}$ (left) and $\Sigma{{\rm SFR_{AGN, H\alpha + \text{[O\:{\sc iii}]}}}}$ (right) across all 22,084 spaxels. The identity line $y=x$ is shown as continuous, and the filled circles represent the mean value with a standard deviation of 20 linearly spaced bins over $\Sigma{\rm SFR_{\rm AGN}}$. The top panels show the mean fractional difference of $\Sigma{\rm SFR}$, defined as $f_{\rm frac}=\langle\log\:\Sigma{\rm SFR_{R21}} -\log\:\Sigma{\rm SFR_{AGN}}\rangle/\sigma_{\Sigma{\rm SFR_{R21}}}$, where $\sigma_{\Sigma{\rm SFR_{R21}}}$ is the standard deviation of ${\log\:\Sigma{\rm SFR_{R21}}}$ within each bin. }
 \label{fig:plane}
\end{figure*}

In the bottom left panel of Fig. \ref{fig:plane} we present the comparison of $\Sigma{\text{SFR}_{\text{R21}}}$ with $\Sigma{\text{SFR}_{\rm AGN,H\alpha}}$. The solid line shows the 1:1 relation, while the black circles correspond to the mean $\Sigma{\text{SFR}_{\text{R21}}}$ values calculated within bins of  $\Sigma{\text{SFR}_{\rm AGN,H\alpha}}$ and the error bars correspond to their standard deviations. As expected, the derived SFR densities based on the H$\alpha$ emission are consistent with the observed values, derived from the stellar population synthesis. In the top left panel, we show the mean fractional difference of $\Sigma{\text{SFR}}$, defined as $f_{\rm frac}=\langle\log\:\Sigma{\rm SFR_{R21}} -\log\:\Sigma{\rm SFR_{AGN}}\rangle/\sigma_{\Sigma{\rm SFR_{R21}}}$, where $\sigma_{\Sigma{\rm SFR_{R21}}}$ is the standard deviation of ${\log\:\Sigma{\rm SFR_{R21}}}$ within each bin. The mean values of $f_{\rm frac}$ are very close to zero for the entire range of observed SFR densities, with a slight systematic deviation for the bins at both ends. The standard deviation in $f_{\rm frac}$ is close to one for most locations, meaning that the estimated  $\Sigma{\text{SFR}_{\rm AGN,H\alpha}}$ in our sample are in agreement with $\Sigma{\rm SFR_{R21}}$, considering its scatter within each bin. 

%As observed the derived SFR densities based on the H$\alpha$ emission are consistent with the observed values, derived from the stellar population synthesis, with a root mean square error (RMSE) of 0.45 dex. 

\begin{figure}
\includegraphics[width=0.48\textwidth]{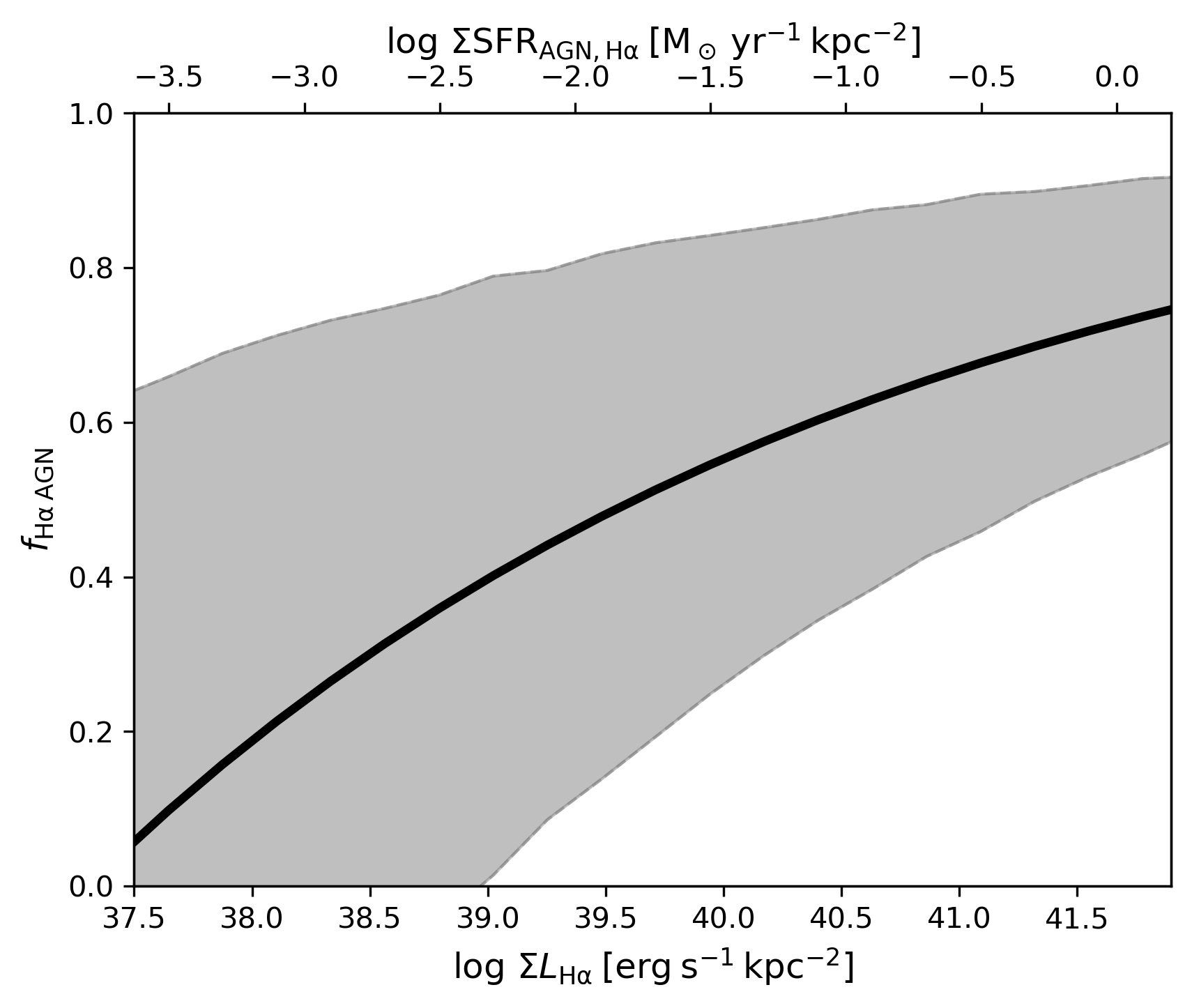}
\caption{Fraction of the H$\alpha$ luminosity ($f_{\rm H\alpha\:AGN}$) due to the AGN emission relative to the observed H$\alpha$ luminosity (AGN+SF contributions) vs. $\log\:\Sigma{\rm L_{H\alpha}}$ (bottom axis) and  $\Sigma{\rm SFR_{\rm AGN,H\alpha}}$ (top axis). The gray region represents the uncertainties estimated from 10\:000 bootstrap realizations, considering the uncertainties in the line coefficients in Eq.~\ref{ajuste_total}. } 
\label{fig:fraction}
\end{figure}

By comparing Eq.~\ref{ajuste_total} with the \citet{kennicutt1998} relation, which is given by 

\begin{equation}
    \Sigma{\rm SFR} = 7.9\times10^{-42} ~\Sigma{\rm L_{H\alpha, SF}} \ ,
\end{equation} we can estimate the fraction of the H$\alpha$ luminosity that is due to the AGN  ($f_{\rm H\alpha\:AGN}$), as Eq.~\ref{ajuste_total} uses the observed luminosity given by the sum of the AGN and SF contributions. This leads to

\begin{align}
    \log{(\Sigma{ \rm L_{H\alpha, AGN}})} & = 10^{[( \ \log{(7.9 \times 10^{-42} \ \Sigma{ \rm L_{H\alpha, SF}})} \ + \ 36.273) / 0.87]}
    \\ & - \log{(\Sigma{\rm L_{H_\alpha, SF}})}, \nonumber
\end{align} then the fraction is given by 

\begin{equation}
    f_{\rm H\alpha\:AGN} =  \frac{\log{(\Sigma{ \rm L_{H\alpha,AGN}})}}{\log{(\Sigma{ \rm L_{H\alpha,AGN}})} + \log{(\Sigma{\rm L_{H_\alpha,SF}})}}.
\end{equation}
Figure~\ref{fig:fraction} shows the plot of  $f_{\rm H\alpha\:AGN}$ vs. $\log\:\Sigma{\rm L_{H\alpha}}$ (bottom axis) and  $\Sigma{\rm SFR_{\rm AGN,H\alpha}}$ (top axis), for the observed H$\alpha$ luminosity range of our sample. For the lowest luminosity spaxels, $f_{\rm H\alpha\:AGN}$ is close to zero, increasing with the H$\alpha$ luminosity to up to 80\,\% for the brightest spaxels. 

In addition, we note that the [O\,{\sc iii}]$\lambda$5007/H$\beta$ line ratio increases with the distance from the \citet{kennicutt1998} relation. We compute the  median [O\,{\sc iii}]$\lambda$5007/H$\beta$ values within bins of 0.5 dex from the \citet{kennicutt1998} relation. These values range from log\:[O\,{\sc iii}]$\lambda$5007/H$\beta$=0.30 for the bin centred at 0.75 dex above the \citet{kennicutt1998} relation to log\:[O\,{\sc iii}]$\lambda$5007/H$\beta$=0.44 for the bin centred at 2.25 dex below the relation. This result is expected, as AGN with larger values [O\,{\sc iii}]$\lambda$5007/H$\beta$ are also expected to be more luminous, thus having a larger fraction of the H$\alpha$ luminosity produced by the AGN, relative to the total H$\alpha$ luminosity, including the AGN and SF contributions.  This trend is further illustrated in Fig.~\ref{fig:ajustereta}, where, for a constant H$\alpha$ luminosity, we observe higher [O\,{\sc iii}]$\lambda$5007 luminosities associated with lower $\Sigma{\rm SFR_{\rm R21}}$ values.

 The [O\,{\sc iii}]$\lambda$5007 is a better tracer of the AGN luminosity than the H$\alpha$, as it traces higher ionization gas emission. In order to test the dependence of $\Sigma{\text{SFR}_{\text{R21}}}$ with the AGN luminosity, we fitted the data by a plane given by $\log (\Sigma{\rm SFR_{AGN,H\alpha+[OIII]}})=A\:\log(\Sigma{\rm L_{H\alpha}})+B\:\log(\Sigma{\rm L_{[OIII])}}) + C$, with $A, B$ and $C$ being the parameters to be determined. The best-fit equation is
 
\begin{align}\label{eq:OIII}
    \log (\Sigma{\rm SFR_{AGN,H\alpha+[OIII]}})  & = \ (1.211 \pm 0.014) \:\log(\Sigma{\rm L_{H\alpha}}) \nonumber \\ & - \ (0.327 \pm 0.012) \:\log(\Sigma{\rm L_{[OIII])}}) \nonumber \\ & - \ (36.774 \pm 0.250),
\end{align} using the same units as those in Eq.~\ref{ajuste_total}.
%that can be utilized to determine the SFR densities in regions dominated by AGN ionization, based on the luminosity of H$\alpha$ and [O\:{\sc iii}].

In the bottom right panel of Fig. \ref{fig:plane}, we illustrate the comparison between $\Sigma{\text{SFR}_{\text{R21}}}$ and $\Sigma{\rm SFR_{AGN,H\alpha+[OIII]}}$ given by Eq.~\ref{eq:OIII}. The solid line represents the 1:1 relationship, while the black circles correspond to the mean $\Sigma{\text{SFR}_{\text{R21}}}$ values calculated within bins of $\Sigma{\rm SFR_{AGN,H\alpha+[OIII]}}$, with error bars indicating their standard deviations. It is worth noting that the derived SFR densities based on the H$\alpha$ and [O\:{\sc iii}] emission lines are consistent with the observed values obtained from stellar population ones. In the top right panel, we present the mean fractional difference of $\Sigma{\text{SFR}}$, showing that the estimated values of $\Sigma{\rm SFR_{AGN,H\alpha+[OIII]}}$ are consistent with the $\Sigma{\text{SFR}_{\text{R21}}}$, considering the observed scatter in the latter.

%with a RMSE of 0.44 dex, slightly lower than the scatter found by using only H$\alpha$. Indeed, the scatter in the relation between $\Sigma{\text{SFR}_{\text{R21}}}$ and $\Sigma{\rm SFR_{AGN}}$ is dominated by the intrinsic scatter in the SFR density based on the stellar population synthesis. This can be seen from the standard deviation of $\Sigma{\text{SFR}_{\text{R21}}}$ within the luminosity bins shown in Fig.~\ref{fig:ajustereta}, that have a mean value of 0.45 dex.

%\subsection{Dependence with the AGN luminosity}

\section{Discussion}\label{sec:disc}

Unlike star-forming galaxies, for which there are various calibrations based on emission features to determine the star formation rate, such calibrations are very scarce in active galaxies. The star formation rate in AGN hosts can be determined through stellar population synthesis \citep{cid2004, starlight2005, rogerio2009, Rogemar_SP_M1066,rogerio2021, rogerio2023, Diniz17_SP,nicolas2018, henrique2024}. For example, \citet{rogerio2021} used star-forming spaxels of a sample composed of active and inactive galaxies from Survey MaNGA, with the aim of comparing the SFR from the \cite{kennicutt1998} relation, with that obtained by the stellar population synthesis carried out by {\sc starlight}. The primary result of their study, for star-forming dominated spaxels in AGN hosts, is represented by Eq. \ref{eq_riffel}. Besides being a time-consuming method, such high S/N spectra are not readily available for most emission-line galaxies. 

Another SFR indicator that can be used in AGN hosts is the [O\,{\sc ii}]$\lambda$3727 emission line \citep{Kewley04,ho05,Kalfountzou2012,Maddox18,Zhuang19}. This line is not excited in both the broad and narrow line regions of the AGN and is strongly ionized by star formation. However, [O\,{\sc ii}] can also be excited in extended emission line regions, contributing with significant flux. To address this issue, the emission line [Ne\,{\sc V}] can be employed as a tracer of AGN contamination, as it is not excited by star formation. One disadvantage of this method is that the [O\:{\sc ii}] emission line is located at the blue end of the optical spectrum, which is not within the observing window of many spectrographs, thus preventing its application to nearby galaxies. Moreover, this region of the spectrum is heavily affected by interstellar extinction, making it difficult to apply the method to more dusty objects. Our calibrations include only strong emission lines, commonly observed in nearby AGN, thus providing an easily applicable method for AGNs in the local universe.

If the \citet{kennicutt1998} relation, originally calibrated for star-forming galaxies, is applied to AGN hosts, we would anticipate an overestimation of the SFR density compared to the real values.
%, provided the AGN does not significantly influence the SFR of the host galaxy. 
This discrepancy is expected due to additional H$\alpha$ emission resulting from gas ionization by the AGN's radiation field. This behaviour is observed in Fig.~\ref{fig:ajustereta}, with the difference between the SFR densities predicted by the  \citet{kennicutt1998} relation relative to the observed ones increasing with H$\alpha$ luminosity. We find that the fraction of H$\alpha$ luminosity attributed to the AGN reaches up to approximately 60\% for the brightest AGNs (Fig.~\ref{fig:fraction}). This result aligns with the mixing sequences outlined in the BPT diagnostic diagram, indicating that the fraction of H$\alpha$ luminosity due to the AGN  is at least 25\% in optically selected active galaxies \citep{mixingI, mixingII}. Similar results are observed with far-infrared data, where the AGN contribution to the infrared luminosity grows with increasing AGN luminosity \citep[e.g.][]{Symeonidis10,Symeonidis16,Symeonidis22,Diaz-Santos17}. In luminous QSOs, this AGN contribution frequently surpasses the stellar-powered emission across wavelengths from optical to submillimeter \citep[e.g.][]{Symeonidis16}. This emphasizes the necessity of correcting for the AGN contribution when calculating the SFR using the H$\alpha$ emission line. 

%\begin{figure*}
% \includegraphics[width=\textwidth]{fig_SFRs.png}
% \caption{Comparison of $\log$ SFR$_{R21}$ with $\log$ SFR$_{AGN}$ involving 22,084 spaxels of 293 active galaxies. The orange circles represent the mean value with standard deviation of 20 linearly spaced bins over $\log$ SFR$_{AGN}$,  while the dotted line indicate the identity $y=x$. The top panel represent the residual of the relation we found. }
% \label{fig:residuals}
%\end{figure*}

\section{Summary}\label{sec:conclusion}

In this study, we evaluate the use of strong optical emission lines to estimate the $\Sigma{\text{SFR}}$ in circumnuclear regions of galaxies dominated by AGN ionization, using the H$\alpha$ and [O\,{\sc iii}]$\lambda$5007 emission line luminosities. To achieve this, we carefully selected 22,084 spaxels with high signal-to-noise ratio and a high quality fit in the population synthesis from a sample of 293 active galaxies observed in the MaNGA survey. In addition, the emission-line ratios place these spaxels in the AGN region of both the BPT and WHAN diagrams. 

We then compare the $\Sigma{\rm {SFR_{R21}}}$ obtained from the stellar population synthesis over the last 20 Myr, with the luminosity of the H$\alpha$ emission line ($L_{\rm H\alpha}$). From this comparison, we derive Eq. \ref{ajuste_total}, which enables the determination of the SFR surface densities in AGN hosts using only the H$\alpha$ emission line. To investigate the dependence of $\Sigma{\rm {SFR_{R21}}}$ with the AGN luminosity, provided by the [O\,{\sc iii}] emission line luminosity ($L_{\text{[O\,{\sc iii}]}}$), we fitted a plane with $L_{\rm H \alpha}$ and $L_{\text{[O\,{\sc iii}]}}$ as variables. By comparing this plane with $\Sigma{\rm {SFR_{R21}}}$, we derived the Eq. \ref{eq:OIII}, which allows for the calculation of SFR surface densities in AGN regions using both emission line luminosities.  

Our results provide a new method for determining the SFR in active galaxies by taking into account the ionization of the gas by the AGN. In particular, the advantage of this method is its simplicity, as it does not require the use of stellar population synthesis, only information from the emission lines in the optical spectrum, which is readily available. Furthermore, this enables further studies aimed at understanding the role of AGN in the star formation processes within their host galaxies.

Finally, it is important to mention that the use of the calibrations presented here can be applied in the study of galaxy samples, and their use for individual objects is not recommended, due to the observed scatter in the SFR densities based on the stellar population synthesis. That introduces an additional uncertainty of 0.45 dex in $\Sigma{\rm SFR}$, as estimated from the root mean square error between the values obtained from our calibrations and the synthesis based ones. This scatter should be added quadratically to the uncertainties in $\Sigma{\rm SFR}$ obtained from the optical emission lines. Furthermore, we do not recommend using the calibrations on spectra integrated over the entire galaxy, as our goal is to establish a calibration for determining the SFR in AGN-dominated regions. Integrating the light from the whole galaxy would contaminate the AGN emission with contributions from other ionization sources.

\section*{Acknowledgements}
We thank an anonymous referee for their valuable comments, which helped improve this paper.
This work was partially supported by the Conselho Nacional de Desenvolvimento Científico e Tecnológico (CNPq) and Coordenação de Aperfeiçoamento de Pessoal de Nível Superior (CAPES). MSZM acknowledges financial support from CAPES (Finance Code 001). RAR acknowledges the support from CNPq (Proj. 303450/2022-3, 403398/2023-1, \& 441722/2023-7), Funda\c c\~ao de Amparo \`a pesquisa do Estado do Rio Grande do Sul (FAPERGS; Proj. 21/2551-0002018-0), and CAPES  (Proj. 88887.894973/2023-00).  GSI acknowledges financial support from the Fundação de Amparo à Pesquisa do Estado de São Paulo (FAPESP), under Projects 2022/11799-9 and 2024/02487-9. ACK thanks FAPESP for the support grant 2020/16416-5 and CNPq. MT acknowledges support from CNPq, under Project 312541/2021-0.

SDSS is managed by the Astrophysical Research Consortium for the Participating Institutions of the SDSS Collaboration including the Brazilian Participation Group, the Carnegie Institution for Science, Carnegie Mellon University, the Chilean Participation Group, the French Participation Group, Harvard-Smithsonian Center for Astrophysics, Instituto de Astrofisica de Canarias, The Johns Hopkins University, Kavli Institute for the Physics and Mathematics of the Universe (IPMU) / University of Tokyo, the Korean Participation Group, Lawrence Berkeley National Laboratory, Leibniz Institut f\"ur Astrophysik Potsdam (AIP), Max-Planck-Institut f\"ur Astronomie (MPIA Heidelberg), Max-Planck-Institut f\"ur Astrophysik (MPA Garching), Max-Planck-Institut f\"ur Extraterrestrische Physik (MPE), National Astronomical Observatories of China, New Mexico State University, New York University, University of Notre Dame, Observat\' orio Nacional / MCTI, The Ohio State University, Pennsylvania State University, Shanghai Astronomical Observatory, United Kingdom Participation Group, Universidad Nacional Aut\'onoma de M\'exico, University of Arizona, University of Colorado Boulder, University of Oxford, University of Portsmouth, University of Utah, University of Virginia, University of Washington, University of Wisconsin, Vanderbilt University, and Yale University.

This research made use of Astropy,\footnote{http://www.astropy.org} a community-developed core Python package for Astronomy \citep{AstropyCollaboration+13,AstropyCollaboration+18}. 

%%%%%%%%%%%%%%%%%%%%%%%%%%%%%%%%%%%%%%%%%%%%%%%%%%
\section*{Data Availability}

The data underlying this paper are available under SDSS collaboration rules, using measurements from \citet{rogerio2023}, which are available at \hyperlink{https://manga.linea.org.br}{https://manga.linea.org.br}.

%\begin{figure*}
%\includegraphics[width=0.8\textwidth]{extra/fig_calib_teste_SF.png}
%\includegraphics[width=0.8\textwidth]{extra/fig_calib_teste.png}
%\caption{\textcolor{red}{Rogemar: ***Extra: Dependencia da densidade de SFR com a contribuicao da FC. Nos SF não incluimos spaxels com FC > 0 para obter a nova calibração entre SFR(gas) e SFR(estrelas). Ja para os AGNs, nao sei se temos muito o que fazer, mas o número de pontos nos bins mais discrepantes eh bem pequeno.***}} 
%\end{figure*}



\bibliographystyle{mnras}
\bibliography{example} % if your bibtex file is called example.bib

%\begin{thebibliography}{99}
%\end{thebibliography}

%%%%%%%%%%%%%%%%%%%%%%%%%%%%%%%%%%%%%%%%%%%%%%%%%%

%%%%%%%%%%%%%%%%% APPENDICES %%%%%%%%%%%%%%%%%%%%%

%\appendix



%%%%%%%%%%%%%%%%%%%%%%%%%%%%%%%%%%%%%%%%%%%%%%%%%%

% Don't change these lines
\bsp	% typesetting comment
\label{lastpage}
\end{document}